\documentstyle[preprint, floats, aps]{revtex}

\begin{document}

\preprint{CALT-68-2055}

\draft

\title{A Class of Quantum Error-Correcting Codes Saturating the Quantum 
Hamming Bound}
\author{Daniel Gottesman\thanks{gottesma@theory.caltech.edu}}
\address{California Institute of Technology, Pasadena, CA 91125}
\maketitle

\begin{abstract}
I develop methods for analyzing quantum error-correcting codes, 
and use these methods to construct an infinite class of codes saturating the 
quantum Hamming bound.  These codes encode $k=n-j-2$ qubits in $n=2^j$ 
qubits and correct $t=1$ error.
\end{abstract}


\section{Introduction}
\label{sec-intro}

Since Shor~\cite{shor1} showed that it was possible to create quantum 
error-correcting codes, there has been a great deal of work on trying to 
create efficient codes.  Calderbank and Shor~\cite{shor2} and 
Steane~\cite{steane} demonstrated a method of converting certain classical 
error-correcting codes into quantum ones, and 
Laflamme~et~al.~\cite{laflamme} and Bennett~et~al.~\cite{bennett1}
produced codes to correct one error that encode 1 qubit in 5 qubits.

Suppose we want to encode $k$ qubits in $n$ qubits.  The space of code words
is then some $2^k$-dimensional subspace of the full $2^n$-dimensional
Hilbert space.  The encodings $|\psi_i \rangle$ of the original $2^k$ basis
states form a basis for the space of code words.  When a coherent error 
occurs, the code states are altered by some linear transformation $M$:
\begin{equation}
|{\psi_i}\rangle \longmapsto M |{\psi_i}\rangle.
\end{equation}
We do not require that $M$ be unitary, which will allow us to also correct
incoherent errors.

Typically, we only consider the possibility of errors that act on no more
than $t$ qubits.  An error that acts non-trivially on exactly $t$ qubits
will be said to have {\em length} $t$.  An error of length~1 only 
acts on a 2-dimensional Hilbert space, so the space of 1-qubit errors is
${\cal M}_2$, the space of $2 \times 2$ matrices.

An error-correction process can be modeled by a unitary linear
transformation that entangles the erroneous states $M |{\psi_i}\rangle$ with
an ancilla $|{A}\rangle$ and transforms the combination to a corrected state
\begin{equation}
\left(M |{\psi_i}\rangle \right) \otimes |{A}\rangle \longmapsto |{\psi_i}
\rangle \otimes |{A_{M}}\rangle.
\end{equation}
Note that the map $M \mapsto |{A_M}\rangle$ must be linear, but not necessarily
one-to-one.  If the map is injective, I will call the code {\em non-degenerate},
and if it is not, I will call the code {\em degenerate}.  A degenerate code
has linearly independent matrices that act in a linearly dependent way
on the code words, while in a non-degenerate code, all of the errors 
acting on the code words produce linearly independent states.
Note that Shor's original code~\cite{shor1} is a degenerate code (phase errors
within a group of 3~qubits act the same way), while the $k=1, n=5$
codes~\cite{laflamme,bennett1} are non-degenerate.

At this point, we can measure the ancilla preparatory to restoring it to its
original state without disturbing the states $|{\psi_i}\rangle$.  This process
will correct the error even if the original state is a superposition of the
basis states:
\begin{equation}
\left(M \sum_{i=1}^{2^k} c_i |{\psi_i}\rangle \right) \otimes |{A}\rangle
\longmapsto
\left( \sum_{i=1}^{2^k} c_i |{\psi_i}\rangle \right) \otimes |{A_M}\rangle
\end{equation}

An incoherent error can be modeled as an ensemble of coherent errors.  Since
the above process corrects all coherent errors, it will therefore also correct
incoherent errors.  After the ancilla is measured and restored to its original
state, the system will once again be in a pure state.  Sufficient and necessary
conditions for the system to form a quantum error-correcting code are given
in \cite{bennett1} and~\cite{laflamme2}.  While errors acting on different
code words must produce orthogonal results, different errors acting on the
same code word can produce non-orthogonal states, even in the non-degenerate
case.

We can use the definition of non-degenerate quantum error-correcting
codes to derive the quantum Hamming bound~\cite{ekert} on their possible
efficiency.  It is not known whether the quantum Hamming bound applies to
degenerate codes, although some recent evidence suggests that it does
not~\cite{shor3,lloyd}.  However, the breeding and hashing protocol presented
by Shor and Smolin~\cite{shor3} and the random matrix encodings mentioned by 
Lloyd~\cite{lloyd} do not give a 100\% chance of successful decoding, even
if only a fixed finite number of errors occur.  There are no known degenerate
codes that guarantee success that violate the quantum Hamming bound.
I show in appendix~\ref{sec-degenerate} that a certain class of degenerate
codes to correct 1 error are, in fact, limited by the quantum Hamming bound.
The question for fully general degenerate codes remains open, although
Knill and Laflamme~\cite{laflamme2} showed that at least 5~qubits are
necessary to correct 1 error.  Below, I will assume the code is non-degenerate.

Since there are 3 possible non-trivial 1-qubit errors, the number of 
possible errors $M$ of length $l$ on an $n$-qubit code is $3^l \left( 
\begin{array}{c} n \\ l \end{array} \right)$.  Each of the states $M
|{\psi_i}\rangle$ must be linearly independent, and all of these different
errors must fit into the $2^n$-dimensional Hilbert space of the $n$ qubits.
Thus, for a code that can correct up to $t$ errors,
\begin{equation}
2^k \sum_{l=0}^t 3^l \left( \begin{array}{c} n \\ l \end{array} \right) \leq 
2^n.
\label{eq:QHB}
\end {equation}
For large $n$, this becomes
\begin{equation}
\frac{k}{n} \leq 1 - \frac{t}{n} \log_2 3 - H(t/n),
\end{equation}
where $H(x) = -x \log_2 x - (1-x) \log_2 (1-x)$.

It is an interesting question whether it is generally possible to attain this 
bound, or whether some more restrictive upper bound holds.  Breeding and
hashing methods~\cite{bennett2,bennett1} can asymptotically saturate the
quantum Hamming bound for large blocks, but have a small but non-zero
probability of failure, even for only one error.  For $t=1$ and $k=1$, the
quantum Hamming bound~(\ref{eq:QHB}) implies $n \geq 5$, so the known 5-qubit
code does saturate the bound.  Below, in section~\ref{sec-code}, I will give a
class of codes saturating the bound for $t=1$ and $n=2^j$ (so $k=n-j-2$).  For
large $n$, the efficiency $k/n$ of
these codes approaches 1.  In this sense, they are the analog of the classical
Hamming codes.  To aid in the construction, in section~\ref{sec-tools}, I will
present some methods for analyzing quantum error-correcting codes.  The
method I present of using code stabilizers to describe codes is also given,
using slightly different language, in \cite{shor4}.

Throughout this paper, I will assume the basis of ${\cal M}_2$ is
\begin{equation}
I = \left( \begin{array}{cc} 1 & 0 \\ 0 & 1 \end{array} \right),\ 
X = \left( \begin{array}{cc} 0 & 1 \\ 1 & 0 \end{array} \right),\ 
Y = \left( \begin{array}{cc} 0 & -1 \\ 1 & 0 \end{array} \right),\ 
Z = \left( \begin{array}{cc} 1 & 0 \\ 0 & -1 \end{array} \right).
\end{equation}
Some of the results will hold for other bases, but many will not.  This basis 
has two important properties: all of the matrices either commute or 
anticommute, and $X^2 = -Y^2 = Z^2 = I$.

\section{Code Stabilizers}
\label{sec-tools}

Suppose we have an $n$-qubit system.  Let us write the matrices $X$, $Y$, 
and $Z$ as $X_i$, $Y_i$, and $Z_i$ when they act on the $i$th qubit.  
Let ${\cal G}$ be the group generated by all $3n$ of these
matrices.\footnote{For $n=1$, ${\cal G}$ is just $D_4$, the symmetry 
group of a square.  For larger
$n$, ${\cal G}$ is $(D_4)^n / ({\bf Z}_2)^{n-1}$.}  Since 
$(X_i)^2 = (Z_i)^2 = I$ and $Y_i = Z_i X_i = -X_i Z_i$, ${\cal G}$ has order 
$2^{2n+1}$ (for each~$i$, we can have $I$, $X_i$, $Y_i$, or $Z_i$, plus a 
possible overall factor of~-1).  The group ${\cal G}$ has a few other useful 
features: every element in ${\cal G}$ squares to $\pm 1$ and if $A, B \in 
{\cal G}$, then either $[A,B] = 0$ or $\{A,B\} = 0$.

The code words of the quantum error-correcting code span a subspace $T$ 
of the Hilbert space.  The group ${\cal G}$ acts on the vectors in $T$.  Let 
${\cal H}$ be the stabilizer of $T$ --- i.e., 
\begin{equation} 
{\cal H} = \{M \in {\cal G} \ {\rm s.t.}\ M |{\psi}\rangle = |{\psi}\rangle
\ \forall \, |{\psi}\rangle \in T \}.
\end{equation}
Now suppose $E \in {\cal G}$ and $\exists \, M \in {\cal H}$ s.t. $\{E, 
M\}=0$.  Then $\forall \, |{\psi}\rangle, |{\phi}\rangle \in T$,
\begin{equation}
\langle{\phi}| E |{\psi}\rangle = \langle{\phi}| EM |{\psi}\rangle = - \langle
{\phi}| ME |{\psi}\rangle = - \langle{\phi}| E |{\psi}\rangle
\end{equation}
so $\langle{\phi}| E |{\psi}\rangle = 0$.

The implications of this are profound.  Suppose $E$ and $F$ are two errors, 
both of length $t$ or less.  Then $E |{\psi}\rangle$ and $F |{\phi}\rangle$ are 
orthogonal for all $|{\psi}\rangle, |{\phi}\rangle \in T$ whenever 
$F^{\dagger}E$ anticommutes with anything in ${\cal H}$.  This is the 
requirement for a non-degenerate code, so to find such a code, we just 
need to pick $T$ and corresponding ${\cal H}$ so that every non-trivial matrix
in ${\cal G}$ of length less than or equal to $2t$ anticommutes with some 
member of ${\cal H}$.

It is unclear whether every quantum error-correcting code in the $X$, $Y$,
$Z$ basis can be completely described by its stabilizer ${\cal H}$.  
Certainly, a large class of codes can be described in this way, and I do not
know of any quantum error-correcting codes that cannot be so described.

Given $T$, we can figure out ${\cal H}$, but it will be much easier to find
codes using the above property if we can pick ${\cal H}$ and deduce a space $T$
of code words.  First I will discuss what properties ${\cal H}$ must have
in order for it to be the stabilizer of a space $T$, then I will discuss how
to choose ${\cal H}$ so that the matrices of length $2t$ or less anticommute
with one of its elements.

Clearly, ${\cal H}$ must be a subgroup of ${\cal G}$.  Also, if $M \in 
{\cal H}$, then $M^2 |{\psi}\rangle = M |{\psi}\rangle = |{\psi}\rangle$ for
$|{\psi}\rangle \in T$, so $M$ cannot square to -1.  Finally, if $M,N \in {\cal
H}$, then
\begin{eqnarray}
MN |{\psi}\rangle & = & |{\psi}\rangle \\
NM |{\psi}\rangle & = & |{\psi}\rangle \\
{[M,N]} |{\psi}\rangle & = & 0
\end{eqnarray}
If $\{M,N\} = 0$, then $[M,N] = 2MN$, but $M$ and $N$ are unitary, and 
cannot have 0 eigenvalues.  Thus, $[M,N]=0$, and ${\cal H}$ must be abelian.

Thus, ${\cal H}$ must be abelian and every element of ${\cal H}$ must square
to 1, so ${\cal H}$ is isomorphic to $({\bf Z}_2)^a$ for some $a$.  It turns 
out that these are sufficient conditions for there to exist non-trivial $T$ 
with stabilizer ${\cal H}$, as long as ${\cal H}$ is not too big.  The largest
subspace $T$ with stabilizer ${\cal H}$ will have dimension $2^{n-a}$.  
To show this, I will give an algorithm for constructing a basis for $T$.
Intuitively, it is unsurprising that this should be the dimension of $T$, 
since each generator of ${\cal H}$ has eigenvalues $\pm 1$ and splits  
the Hilbert space in half.

Consider a state that can be written as a tensor product of $0$s and $1$s.  
This sort of state is analogous to one word of a classical code, so I will
call it a {\em quasi-classical} state.  Sometimes I will distinguish between
quasi-classical states that differ by a phase and sometimes I will not.  
Now, given a quasi-classical state $|{\phi}\rangle$, then 
\begin{equation}
|{\psi}\rangle = \sum_{M \in {\cal H}} M |{\phi}\rangle
\end{equation}
is in $T$,\footnote{In fact, $|{\phi}\rangle$ does not need to be a
quasi-classical state for $|{\psi}\rangle$ to be in $T$.  Any state will do, 
but it is easiest to use quasi-classical states.} since applying an element 
of ${\cal H}$ to it will just rearrange the sum.  I will call $|{\phi}\rangle$
the {\em seed} of the code word $|{\psi}\rangle$.  By the same argument, if 
$M \in {\cal H}$, $M |{\phi}\rangle$ acts as the seed for the same 
quantum code word as $|{\phi}\rangle$.  Not every possible seed will 
produce a non-zero code word.  For instance, suppose ${\cal H} = 
\{I, Z_1 Z_2\}$ and we use $|{01}\rangle$ as our seed.  Then 
$|{\psi}\rangle = I |{01}\rangle + Z_1 Z_2 |{01}\rangle = 0$.

To find elements of $T$, we try quasi-classical states until we get one that
produces non-zero $|{\psi}\rangle$, call it $|{\psi_1}\rangle$.  I will show
later that such a state will always exist.  We can write $|{\psi_1}\rangle$
as a sum of quasi-classical states, any of which could act as its seed.  Pick 
a quasi-classical state that does not appear in $|{\psi_1}\rangle$ and does not
produce 0, and use it as the seed for a second state $|{\psi_2}\rangle$.  
Continue this process for all possible quasi-classical states.  The states 
$|{\psi_i}\rangle$ will then form a basis for $T$.  None of them share a
quasi-classical state.

To see that $\{|{\psi_i}\rangle\}$ is a basis, imagine building up the elements 
of ${\cal H}$ by adding generators one by one.  Suppose ${\cal H} = \langle 
M_1, M_2, \ldots M_a \rangle$ (i.e., ${\cal H}$ is generated by $M_1$ through
$M_a$).  Let ${\cal H}_r$ be the group generated by $M_1$ through $M_r$, and
look at the set $S_r$ of quasi-classical states produced by acting with the 
elements of ${\cal H}_r$ on some given quasi-classical seed 
$|{\phi}\rangle$.  The phases of these quasi-classical states will matter.
The next generator $M_{r+1}$ can do one of three things: 
\begin{enumerate}
\item it can map the seed to some new quasi-classical state not in $S_r$, 
\item it can map the seed to plus or minus itself, or
\item it can map the seed to plus or minus times some state in $S_r$ other than
the seed.
\end{enumerate}
I will call a generator that satisfies case~1 a type~1 generator, and so on.

In the first case, all of the elements of ${\cal H}_{r+1} - {\cal H}_r$ will 
also map the seed outside of $S_r$: If $N \in {\cal H}_{r+1} - {\cal H}_r$, then
$N = M M_{r+1}$ for some $M \in {\cal H}_r$.  Then if $\pm N |{\phi}\rangle
\in S_r$, $N |{\phi}\rangle = \pm M' |{\phi}\rangle$ for some $M' \in 
{\cal H}_r$.  Then $M_{r+1} |{\phi}\rangle = \pm M^{-1} M' |{\phi}\rangle \in
S_r$, which contradicts the assumption.  Thus, $S = S_a$ will always have size
$2^b$, where $b$ is the number of type~1 generators.

In the second case, the new generator must act on each qubit as the 
identity~$I$, as $-I$, or as $Z_i$, so type~2 generators
can be written as the product of $Z$'s.  In principle, a type~2 generator could
be -1 times the product of $Z$'s, but the factor of -1 slightly complicates
the process of picking seeds, so for simplicity I will assume it is not present.
The method of choosing ${\cal H}$ that I give below will always create 
generators without such factors of -1.

In the third case, when $|{\phi '}\rangle = \pm M_{r+1} |{\phi}\rangle$ is
already in $S_r$, then there exists $N \in {\cal H}_r$ with $N |{\phi}\rangle =
|{\phi '}\rangle$.
We can then use $N^{-1}M_{r+1}$ as a new generator instead of $M_{r+1}$, and
since $N^{-1}M_{r+1} |{\phi}\rangle = \pm |{\phi}\rangle$, we are back to case
two.  After adding all of the generators, changing any of type~3 into type~2, 
we are left with $b$ generators of type~1 and $a-b$ generators of type~2.

If one of the type~2 generators~$M_i$ gives a factor of -1 acting on the seed, 
the final state is~$0$:
\begin{equation}
\sum_{M \in {\cal H}} M |{\phi}\rangle = \left(\sum_{M \in {\cal H}} M \right)
M_i |{\phi}\rangle = - \sum_{M \in {\cal H}} M |{\phi}\rangle = 0.
\end{equation}
Otherwise $|{\psi}\rangle$ is non-zero.  We can simplify the computation of
$|{\psi}\rangle$ by only summing over products of the type~1 generators, since
the type~2 generators will only give us additional copies of the same sum.
Then $|{\psi}\rangle$ will be the sum of $2^b$ quasi-classical states (with the
appropriate signs).

Is this classification of generators going to be the same for all possible
seeds?  Anything that is a product of~$Z$'s has all quasi-classical states as
eigenstates, and anything that is not a product of~$Z$'s has no quasi-classical
states as eigenstates.  Thus if a generator is type~2 for one seed, it is
type~2 for all seeds.  Type~1 generators cannot become type~3 generators because
then the matrix $M^{-1}N$ would be type~2 for some states but not others.  
Thus, all of the states $|{\psi_i}\rangle$ are the sum of $2^b$ quasi-classical
states, and $a-b$ of the generators of ${\cal H}$ are the product of~$Z$'s.
Note that this also shows that the classification of generators into type~1 
and type~2 generators does not depend on their order.

Since a seed produces a non-trivial final state if and only if it has an 
eigenvalue of +1 for all of the type~2 generators, all of the states 
$|{\psi_i}\rangle$ live in the joint +1~eigenspace of the $a-b$ type~2
generators, which has dimension $2^{n-(a-b)}$.  We can partition the
quasi-classical basis states of this eigenspace into classes based on the
$|{\psi_i}\rangle$ in which they appear.  Each partition has size $2^b$, so
there are $2^{n-a}$ partitions, proving the claimed dimension of $T$.  The
states $|{\psi_i}\rangle$ form a basis of $T$.

We can simplify the task of finding seeds for a basis of quantum code words.
First, note that $|{\bf 0}\rangle = |{00 \ldots 0}\rangle$ is always in the
+1~eigenspace of any type~2 generator, so it can always provide our first seed. 
Any other quasi-classical seed $|{\phi}\rangle$ can be produced from
$|{\bf{0}}\rangle$
by operating with some $N \in {\cal G}$ that is a product of $X$'s.  For 
$N |{\bf{0}}\rangle$ to act as the seed for a non-trivial state, $N$ must
commute with every type~2 generator in ${\cal H}$: If $M_i$ is a type~2
generator, and $\{N, M_i \} = 0$, then
\begin{equation}
M_i (N |{\bf{0}}\rangle) = - N M_i |{\bf{0}}\rangle = 
- N |{\bf{0}}\rangle.
\end{equation}
But only quasi-classical states which have eigenvalue~+1 give non-trivial
code words, so $N$ must commute with the type~2 generators.
Two such operators $N$ and $N'$ will produce seeds for the same quantum code
word iff they differ by an element of ${\cal H}$ --- i.e., $N^{-1}N' \in 
{\cal H}$.
This provides a test for when two seeds will produce different code words,
and also implies that the product of two operators producing different code
words will also be a new code word.  Thus, we can get a full set of $2^{n-a}$
seeds by taking products of $n-a$ operators $N_1, \ldots, N_{n-a}$.  I will
call the $N_i$ {\em seed generators}.  I do not know of any efficient method
for determining the $N_i$.

Once we have determined the generators $M_i$ of ${\cal H}$ and the seed
generators $N_i$, we can define a unitary transformation to perform the
encoding by
\begin{equation}
|c_1 c_2 \ldots c_k \rangle \longmapsto 
\frac{1}{2^{b/2}} \prod_{M_i\ {\rm type\ 1}} (I + M_i)\ N_1^{c_1} N_2^{c_2}
\ldots N_k^{c_k} |{\bf 0} \rangle.
\end{equation}
However, I do not know of an efficient way to implement this tranformation.

Now I turn to the next question: how can we pick ${\cal H}$ so that all of the 
errors up to length $2t$ anti-commute with some element of it?  Given $M 
\in {\cal G}$, consider the function $f_M : {\cal G} \rightarrow 
{\bf Z}_2$,
\begin{equation}
f_M (N) = \left\{ \begin{array}{ll} 0 & \mbox{if $[M,N]=0$} \\ 1 & \mbox{if 
$\{M,N\}=0$} \end{array} \right.
\end{equation}
Then $f_M$ is a homomorphism.  If ${\cal H} = \langle M_1, M_2, \ldots M_a 
\rangle$, then define a homomorphism $f: {\cal G} \rightarrow ({\bf Z}_2)^a$
by
\begin{equation}
f (N) = \left(f_{M_1}(N), f_{M_2}(N), \ldots f_{M_a} (N) \right).
\end{equation}
Below, I will actually write $f(N)$ as an $a$-bit binary string.  With this 
definition of $f$, $f(N) = 00\ldots0$ iff $N$ commutes with everything in 
${\cal H}$.  We therefore wish to pick ${\cal H}$ so that $f(E)$ is non-zero 
for all $E$ up to length $2t$.  We can write any such $E$ as the product of $F$ 
and $G$, each of length $t$ or less, and $f(E) \neq 0$ iff $f(F) \neq f(G)$.  
Therefore, we need to pick ${\cal H}$ so that $f(F)$ is different for each 
$F$ of length $t$ or less.

We can thus find a quantum error-correcting code by first choosing a 
different $a$-bit binary number for each $X_i$ and $Z_i$.  These numbers 
will be the values of $f(X_i)$ and $f(Z_i)$ for some ${\cal H}$ which we can 
then determine.  We want to pick these binary numbers so that the 
corresponding values of $f(Y_i)$ and errors of length~2 or more (if $t>1$) 
are all different.  While this task is difficult in general, it is tractable 
for $t=1$.  In addition, even if all of the $f(E)$ are different, we still 
need to make sure that ${\cal H}$ fixes a non-trivial space of code words 
$T$ by checking that ${\cal H}$ is abelian and that its elements square to +1.

\section{The Codes}
\label{sec-code}

Now I will use the method described in section~\ref{sec-tools} to construct 
an optimal non-degenerate quantum error-correcting code for $n=2^j$.  
The quantum Hamming bound~(\ref{eq:QHB}) tells us that $k \leq n-j-2$, so 
we take $a=j+2$ and $j \geq 3$.  I will also show explicitly the construction 
for $n=8$.  Steane~\cite{steane2} has found the same $k=3$, $n=8$ code
following inspiration from classical error-correcting codes, and Calderbank
et~al.~\cite{shor4} have found a different $k=3$, $n=8$ code.

We want to pick different $(j+2)$-bit binary numbers for $X_i$ and $Z_i$ 
($i=1 \ldots n$) so that the numbers for $Y_i$, which are given by the 
bitwise XOR of the numbers for $X_i$ and $Z_i$, are also all different.  The
numbers for $n=8$ are shown in table~\ref{fig:binary8}.  
\begin{table}
\begin{tabular}{llllllll}
$X_1$ & 01000 & $X_2$ & 01001 & $X_3$ & 01010 & $X_4$ & 01011 \\
$Z_1$ & 10111 & $Z_2$ & 10000 & $Z_3$ & 10110 & $Z_4$ & 10001 \\
$Y_1$ & 11111 & $Y_2$ & 11001 & $Y_3$ & 11100 & $Y_4$ & 11010 \\
\hline
$X_5$ & 01100 & $X_6$ & 01101 & $X_7$ & 01110 & $X_8$ & 01111 \\
$Z_5$ & 10010 & $Z_6$ & 10101 & $Z_7$ & 10011 & $Z_8$ & 10100 \\
$Y_5$ & 11110 & $Y_6$ & 11000 & $Y_7$ & 11101 & $Y_8$ & 11011 \\
\end{tabular}
\caption{The values of $f(X_i)$, $f(Y_i)$, and $f(Z_i)$ for $n=8$.}
\label{fig:binary8}
\end{table}
In order to distinguish between the $X$'s, the $Y$'s, and the $Z$'s, we will
devote the first two bits to encoding which of the three it is, and the
remaining $j$ bits will encode which qubit $i$ the error acts on (although this
encoding will depend on whether it is an $X$, a $Y$, or a $Z$).

The first two bits are $01$ for an $X$, $10$ for a $Z$, and $11$ for a $Y$, 
as required to make $f$ a homomorphism.  For the $X_i$'s, the last $j$ bits 
will just form the binary number for $i-1$, so $X_1$ is $0100 \ldots 0$, 
and $X_n$ is $0111 \ldots 1$.  The encoding for the last $j$ bits for the 
$Z_i$'s is more complicated.  We cannot use the same pattern, or all of the 
$Y_i$'s would just have all $0$s for the last $j$ bits.  Instead of counting 0, 
1, 2, 3, \ldots, we instead count 0, 0, 1, 1, 2, 2, \ldots.  Writing this in 
binary will not make all of the numbers for the $Z$'s different, so what we 
do instead is to write them in binary and then take the bitwise NOT of one 
of each pair.  This does make all of the $Z$'s different.  We then determine 
what the numbers for $Y_i$ are.

How we pick which member of the pair to invert will determine whether 
all of the numbers for $Y_i$ are different.  For even $j$, we can just take 
the NOT for all odd $i$; but for odd $j$, we must take the NOT for odd $i$ 
when $i \leq 2^{j-1}$ and for even $i$ when $i > 2^{j-1}$.
A general proof that this method will give different numbers for all the 
$Y_i$'s is given in appendix~\ref{sec-different}.

Now that we have the numbers for all of the 1-qubit errors, we need to 
determine the generators $M_1 \ldots M_a$ of ${\cal H}$.  Recall that the first 
digit of the binary numbers corresponds to the first generator.  Since the 
first digit of the number for $X_1$ is $0$, $M_1$ commutes with $X_1$; 
the first digit of the numbers for $Y_1$ and $Z_1$ are both $1$, so $M_1$ 
anticommutes with $Y_1$ and $Z_1$.  Therefore, $M_1$ is $X_1$ times the 
product of matrices which only act on the other qubits.  Similarly, the first 
digit of the number for each $X_i$ is $0$ and the first digits for $Y_i$ and 
$Z_i$ are both $1$, so $M_1=X_1 X_2 \ldots X_n$ (this is true even for 
$j>3$).  Using the same principle, we can work out all of the generators.  

The results for $n=8$ are summarized in table~\ref{fig:group8}.
\begin{table}
\begin{tabular}{|l|cccccccc|}
$M_1$ & $X_1$ & $X_2$ & $X_3$ & $X_4$ & $X_5$ & $X_6$ & $X_7$ & $X_8$ \\
$M_2$ & $Z_1$ & $Z_2$ & $Z_3$ & $Z_4$ & $Z_5$ & $Z_6$ & $Z_7$ & $Z_8$ \\
$M_3$ & $X_1$ & $ I $ & $X_3$ & $ I $ & $Z_5$ & $Y_6$ & $Z_7$ & $Y_8$ \\
$M_4$ & $X_1$ & $ I $ & $Y_3$ & $Z_4$ & $X_5$ & $ I $ & $Y_7$ & $Z_8$ \\
$M_5$ & $X_1$ & $Z_2$ & $ I $ & $Y_4$ & $ I $ & $Y_6$ & $X_7$ & $Z_8$ \\
\hline
$N_1$ & $X_1$ & $X_2$ & $ I $ & $ I $ & $ I $ & $ I $ & $ I $ & $ I $ \\
$N_2$ & $X_1$ & $ I $ & $X_3$ & $ I $ & $ I $ & $ I $ & $ I $ & $ I $ \\
$N_3$ & $X_1$ & $ I $ & $ I $ & $ I $ & $X_5$ & $ I $ & $ I $ & $ I $ \\
\end{tabular}
\caption{The generators of ${\cal H}$ and seed generators for $n=8$.}
\label{fig:group8}
\end{table}
Note that all of these generators square to +1 and that they all commute 
with each other.  A proof of this fact for $j>3$ is given in 
appendix~\ref{sec-commute}.  Thus we have a code that encodes 3~qubits 
in 8~qubits, or more generally $n-j-2$~qubits in $2^j$~qubits.  For these
codes, there is 1~type~2 generator~$M_2$.  The remaining $j+1$ generators
are type~1.

Table~\ref{fig:group8} also gives seed generators for $n=8$.  We can see
immediately that they all commute with $M_2$, the type~2 generator.  It is
less obvious that they all produce seeds for different states, but using them
produces 8~different quantum codes words, listed in table~\ref{fig:words8},
so they do, in fact, form a complete list of seed generators.
\begin{table}
{\tighten
\begin{eqnarray*}
|{\psi_0}\rangle & = & |{00000000}\rangle + |{11111111}\rangle + 
|{10100101}\rangle + |{10101010}\rangle + |{10010110}\rangle + 
|{01011010}\rangle \\
& & \mbox{} + |{01010101}\rangle + |{01101001}\rangle + 
|{00001111}\rangle + |{00110011}\rangle + |{00111100}\rangle \\
& & \mbox{} + |{11110000}\rangle + |{11001100}\rangle + |{11000011}\rangle + 
|{10011001}\rangle + |{01100110}\rangle \\
|{\psi_1}\rangle & = & |{11000000}\rangle + |{00111111}\rangle + 
|{01100101}\rangle + |{01101010}\rangle - |{01010110}\rangle + 
|{10011010}\rangle \\
& & \mbox{} + |{10010101}\rangle - |{10101001}\rangle + |{11001111}\rangle -
|{11110011}\rangle - |{11111100}\rangle \\
& & \mbox{} + |{00110000}\rangle - |{00001100}\rangle - |{00000011}\rangle - 
|{01011001}\rangle - |{10100110}\rangle \\
|{\psi_2}\rangle & = & |{10100000}\rangle + |{01011111}\rangle + 
|{00000101}\rangle - |{00001010}\rangle + |{00110110}\rangle + 
|{11111010}\rangle \\
& & \mbox{} - |{11110101}\rangle + |{11001001}\rangle - |{10101111}\rangle + 
|{10010011}\rangle - |{10011100}\rangle \\
& & \mbox{} - |{01010000}\rangle + |{01101100}\rangle - |{01100011}\rangle - 
|{00111001}\rangle - |{11000110}\rangle \\
|{\psi_3}\rangle & = & |{01100000}\rangle + |{10011111}\rangle + 
|{11000101}\rangle - |{11001010}\rangle - |{11110110}\rangle + 
|{00111010}\rangle \\
& & \mbox{} - |{00110101}\rangle - |{00001001}\rangle - |{01101111}\rangle - 
|{01010011}\rangle + |{01011100}\rangle \\
& & \mbox{} - |{10010000}\rangle - |{10101100}\rangle + |{10100011}\rangle + 
|{11111001}\rangle + |{00000110}\rangle \\
|{\psi_4}\rangle & = & |{10001000}\rangle + |{01110111}\rangle - 
|{00101101}\rangle + |{00100010}\rangle + |{00011110}\rangle - 
|{11010010}\rangle \\
& & \mbox{} + |{11011101}\rangle + |{11100001}\rangle - |{10000111}\rangle - 
|{10111011}\rangle + |{10110100}\rangle \\
& & \mbox{} - |{01111000}\rangle - |{01000100}\rangle + |{01001011}\rangle - 
|{00010001}\rangle - |{11101110}\rangle \\
|{\psi_5}\rangle & = & |{01001000}\rangle + |{10110111}\rangle - 
|{11101101}\rangle + |{11100010}\rangle - |{11011110}\rangle - 
|{00010010}\rangle \\
& & \mbox{} + |{00011101}\rangle - |{00100001}\rangle - |{01000111}\rangle + 
|{01111011}\rangle - |{01110100}\rangle \\
& & \mbox{} - |{10111000}\rangle + |{10000100}\rangle - |{10001011}\rangle + 
|{11010001}\rangle + |{00101110}\rangle \\
|{\psi_6}\rangle & = & |{00101000}\rangle + |{11010111}\rangle - 
|{10001101}\rangle - |{10000010}\rangle + |{10111110}\rangle - 
|{01110010}\rangle \\
& & \mbox{} - |{01111101}\rangle + |{01000001}\rangle + |{00100111}\rangle - 
|{00011011}\rangle - |{00010100}\rangle \\
& & \mbox{} + |{11011000}\rangle - |{11100100}\rangle - |{11101011}\rangle + 
|{10110001}\rangle + |{01001110}\rangle \\
|{\psi_7}\rangle & = & |{11101000}\rangle + |{00010111}\rangle - 
|{01001101}\rangle - |{01000010}\rangle - |{01111110}\rangle - 
|{10110010}\rangle \\
& & \mbox{} - |{10111101}\rangle - |{10000001}\rangle + |{11100111}\rangle + 
|{11011011}\rangle + |{11010100}\rangle \\
& & \mbox{} + |{00011000}\rangle + |{00100100}\rangle + |{00101011}\rangle - 
|{01110001}\rangle - |{10001110}\rangle \\
\end{eqnarray*}}

\caption{The quantum code words for the $n=8$ code.}
\label{fig:words8}
\end{table}
This partly answers the question of how often we can saturate the 
quantum Hamming bound by showing that for 1 error, it can be saturated 
for arbitrarily large $n$.  Although the methods given above may help 
somewhat, finding optimal codes to correct more than one error remains a 
difficult task.

I would like to thank John Preskill for helpful discussions.  This work was 
supported in part by the U.S.~Department of Energy under Grant No. 
DE-FG03-92-ER40701.

\appendix

\section{Proof that certain degenerate codes cannot defeat the quantum Hamming
bound for $\lowercase{t}=1$}
\label{sec-degenerate}

While there is no known proof that degenerate quantum error-correcting codes
cannot beat the quantum Hamming bound for arbitrary $t$ and $n$, I will 
present a proof that codes to correct just 1~error are, in fact, limited 
by that bound, so long as the only source of degeneracies is when
linearly independent error matrices map a code word into a one-dimensional
subspace.  For instance, if three different errors map code words into a 
single two-dimensional subspace, this condition will not generally
be satisfied.

Given a degenerate quantum error-correcting code of this type
that corrects 1 error, we can list a number of conditions that 
describe which errors are degenerate.  I will call these relations {\em
degeneracy conditions.}  As with the stabilizers in
section~\ref{sec-tools}, each independent condition will reduce 
the space of possible code words by a factor of~2.  Note that I am not
requiring that the basis for errors be the $X$, $Y$, $Z$ basis I have used in
the rest of the paper.\footnote{The proof that the dimension of $T$ is
$2^{n-l}$ given in section~\ref{sec-tools} only
works for the $X$, $Y$, $Z$ basis, but for this appendix, I only need the
weaker result that the dimension of $T$ is at least halved by any degeneracy
condition that constrains a qubit unaffected by any of the other degeneracy
conditions.  This should be self-evident.} 

Suppose there are $l$
different degeneracy conditions describing the code.  Each one equates two 
one-qubit errors, so at most $2l$ qubits are affected by the degenerate errors.
The errors on the remaining $n-2l$ qubits must produce mutually orthogonal
states.  There are $3 (n-2l)$ possible errors affecting those qubits.

Furthermore, errors on those qubits commute with the degenerate errors, since
they act on different qubits, so if $M |{\psi_i}\rangle = N |{\psi_i}\rangle$
and $E$ is an error that acts on a qubit unaffected by the degenerate errors,
\begin{equation}
M E |{\psi_i}\rangle = E M |{\psi_i}\rangle = E N |{\psi_i}\rangle = 
N E |{\psi_i}\rangle.
\end{equation}
Thus, the state $E |{\psi_i}\rangle$ still satisfies the same set of
degeneracy conditions.  The space of states that satisfy the given set of
$l$ degeneracy conditions has dimension at most $2^{n-l}$.  To fit
all the states $E |{\psi_i}\rangle$ inside it, if $l \leq n/2$, we must have
\begin{equation}
\left[ 1+3(n-2l) \right] 2^k \leq 2^{n-l},
\label{eq:degQHB}
\end{equation}
or
\begin{equation}
k \leq n - l - \log_2 \left[ 1+3(n-2l) \right] = g(l).
\end{equation}
For $l=0$, this becomes the quantum Hamming bound.  Now,
\begin{equation}
\frac{{\rm d}g}{{\rm d}l} = -1 + \frac{6/\ln 2}{1+3(n-2l)}.
\end{equation}
Therefore $g(l)$ is decreasing for
\begin{equation}
1+3(n-2l) \geq \frac{6}{\ln 2}
\end{equation}
\begin{equation}
l \leq \frac{n}{2} - \left( \frac{1}{\ln 2} - \frac{1}{6} \right).
\end{equation}
Thus, the quantum Hamming bound holds for $l \leq (n-3)/2$.  For $l>(n-3)/2$,
we still have $k \leq n-l < (n+3)/2$.  This automatically satisfies the
quantum Hamming bound for $n \geq 13$ (see table~\ref{fig:QHBvalues}).
\begin{table}
\begin{tabbing}
\rule[-.1in]{.7in}{.01in}\hspace{-.7in}n\hspace{.5in}\= k \\
5 \> 1 \\
6 \> 1 \\
7 \> 2 \\
8 \> 3 \\
9 \> 4 \\
10 \> 5 \\
11 \> 5 \\
12 \> 6 \\
13 \> 7
\end{tabbing}
\caption{The maximum $k$ allowed by the quantum Hamming bound for $n \leq 13$}
\label{fig:QHBvalues}
\end{table}

For $n<13$, $l>(n-3)/2$, we need a different argument.  When $l < n-1$,
there must always be at least one degeneracy condition that relates errors on
two qubits that are unaffected by any other degeneracy conditions.  There are
three possible errors on each qubit, and only one pair of them are going to
produce the same results, so there are still 5 different errors, plus the
possibility of no error.  As above, these errors will remain within the space
that satisfies the other $l-1$ degeneracy conditions, so
\begin{equation}
(1+5) \, 2^k \leq 2^{n-(l-1)},
\end{equation}
or $k \leq n-l+(1-\log_2 6)$ (i.e., $k \leq n-l-2$).  When $l>(n-3)/2$, this
means $k \leq (n-1)/2$.  Applying this condition for $n \leq 12$ restricts
violations of the quantum Hamming bound to $n \leq 6$, specifically: $n=6$ and
$l=2$, and $n=4$ and $l=1$.  For these two cases, we can directly apply 
equation~(\ref{eq:degQHB}) to see that for $n=6$ and $l=2$, $k \leq 1$, in
accordance with the quantum Hamming bound; and for $n=4$ and $l=1$, $k=0$.

Finally, for $l = n-1$, there must be at least one qubit that is only affected
by a single degeneracy condition.  All three errors on this qubit commute with
the other $n-2$ degeneracy conditions, so
\begin{equation}
(1+3) \, 2^k \leq 2^{n-(n-2)}
\end{equation}
Therefore $k=0$, and the quantum Hamming bound holds for any  degenerate
quantum code where linearly independent errors can only map code words
into a one-dimensional subspace.

\section{Proof that the numbers for $Y_{\lowercase{i}}$ are all different}
\label{sec-different}

The construction of the numbers for $X_i$ and $Z_i$ immediately 
demonstrates that they are all different.  However, it is not as clear that all 
of the numbers for the $Y_i$'s, which are determined by the numbers for 
the $X_i$'s and $Z_i$'s, will also be different.  The first two bits just
enforce the requirement that any $Y_i$ is different from an $X$ or a $Z$, so 
I will only consider the last $j$ bits.  All references to bit number in this 
appendix will refer to position within the last $j$ bits, so bit number~``1'' 
is actually bit~3, and bit~``$l$'' is actually bit~$l+2$.

Consider the pictorial representation of the algorithm to pick the errors' 
binary numbers given in table~\ref{fig:numberchart}.
\begin{table}
\begin{tabular}{|c|c|c|c|c|c|c|c|c|c|c|c|c|c|c|c|}
\multicolumn{16}{c}{$X_i$:} \\
\hline
\multicolumn{8}{|c|}{0} & \multicolumn{8}{c|}{1} \\
\hline
\multicolumn{4}{|c|}{0} & \multicolumn{4}{c|}{1} & \multicolumn{4}{c|}{0} & 
\multicolumn{4}{c|}{1} \\
\hline
\multicolumn{2}{|c|}{0} & \multicolumn {2}{c|}{1} & \multicolumn{2}{c|}{0} & 
\multicolumn {2}{c|}{1} & \multicolumn{2}{c|}{0} & \multicolumn {2}{c|}{1} & 
\multicolumn{2}{c|}{0} & \multicolumn {2}{c|}{1} \\
\hline
0 & 1 & 0 & 1 & 0 & 1 & 0 & 1 & 0 & 1 & 0 & 1 & 0 & 1 & 0 & 1 \\
\hline
\multicolumn{16}{c}{\vdots}\\
\multicolumn{16}{c}{} \\
\multicolumn{16}{c}{$Z_i$: Parity XOR} \\
\hline
\multicolumn{8}{|c|}{0} & \multicolumn{8}{c|}{0} \\
\hline
\multicolumn{4}{|c|}{0} & \multicolumn{4}{c|}{0} & \multicolumn{4}{c|}{1} & 
\multicolumn{4}{c|}{1} \\
\hline
\multicolumn{2}{|c|}{0} & \multicolumn {2}{c|}{0} & \multicolumn{2}{c|}{1} & 
\multicolumn {2}{c|}{1} & \multicolumn{2}{c|}{0} & \multicolumn {2}{c|}{0} & 
\multicolumn{2}{c|}{1} & \multicolumn {2}{c|}{1} \\
\hline
0 & 0 & 1 & 1 & 0 & 0 & 1 & 1 & 0 & 0 & 1 & 1 & 0 & 0 & 1 & 1 \\
\hline
\multicolumn{16}{c}{\vdots}\\
\multicolumn{16}{c}{} \\
\multicolumn{16}{c}{$Y_i$: Parity XOR} \\
\hline
\multicolumn{8}{|c|}{0} & \multicolumn{8}{c|}{1} \\
\hline
\multicolumn{4}{|c|}{0} & \multicolumn{4}{c|}{1} & \multicolumn{4}{c|}{1} & 
\multicolumn{4}{c|}{0} \\
\hline
\multicolumn{2}{|c|}{0} & \multicolumn {2}{c|}{1} & \multicolumn{2}{c|}{1} & 
\multicolumn {2}{c|}{0} & \multicolumn{2}{c|}{0} & \multicolumn {2}{c|}{1} & 
\multicolumn{2}{c|}{1} & \multicolumn {2}{c|}{0} \\
\hline
0 & 1 & 1 & 0 & 0 & 1 & 1 & 0 & 0 & 1 & 1 & 0 & 0 & 1 & 1 & 0 \\
\hline
\multicolumn{16}{c}{\vdots }\\
\end{tabular}

\caption{The first 4 bits (of the last $j$) of the numbers for $X_i$, $Y_i$, 
and $Z_i$.  The $p$th row corresponds to the $p$th bit and the columns in the
$p$th row correspond to the possible values for the first $p$ bits of $i$.
For $Y_i$ and $Z_i$, the actual numbers require an additional XOR with the
parity or reverse of the parity of $i$.}
\label{fig:numberchart}
\end{table}
The numbers given for $X_i$ are the actual numbers that appear.  For 
$Y_i$ and $Z_i$, we need to take an XOR with the parity of $i$ (for $j$ even 
or $i \leq 2^{j-1}$), or an XOR with the reverse of the parity of $i$ (for $j$ 
odd and $i>2^{j-1}$).  We can see that before we apply the XOR, the number for 
$Y_i$ encodes $i$ in a unique fashion, since if $i$ and $i'$ first differ in 
the $r$th bit, then the numbers for $Y_i$ and $Y_{i'}$ will also differ in the
$r$th bit.  The only way we could get two of the numbers to be the same would 
be if the XOR operation reverses one of a pair that would normally have 
complementary values in all bits.

Does this ever happen?  Given a number $f(Y_i)$ for $i \leq n/2$, the number
with complementary bits must appear for $i>n/2$, since the first digit 
does not change until then.  The XOR will therefore collapse these two 
numbers into one whenever the parity of the appropriate $i$'s is the same 
(for $j$ odd) or different (for $j$ even).

Pick some bit string starting with 0.  There will be an $i \leq n/2$ such
that $Y_i$ has that number.  Which $i'$ will have the complementary bitstring?
If we take the binary representation of $i$, it will begin with a 0 and the 
binary representation of $i'$ will begin with a 1.\footnote{I am ignoring the
special case of $i=n/2$, which works on the same principle after the first
digit of $i$.}  The next digit of $i$ can 
be either 0 or 1, and from table~\ref{fig:numberchart}, we can see $i'$ will 
have the same value for this digit.  The third digits of $i$ and $i'$ will be 
opposite again.  In general, a 0 in the $r$th~digit of $i$ or $i'$ means the 
two squares relevant to the next digit will read 01, while a 1 in the $r$th 
digit will mean the two squares for the next digit will read 10.  Thus, if $i$ 
and $i'$ agree in the $r$th~digit, they will disagree in the next digit, 
and vice-versa.  Thus, $i$ and $i'$ agree on even-numbered digits and 
disagree on odd-numbered digits.

This means the last digit agrees for $j$ even and disagrees for $j$ 
odd.  Therefore, the XOR will not make $Y_i$ the same as $Y_{i'}$ --- it will 
either reverse both of them or neither of them.  This explains why 
different rules for odd and even $j$ were necessary.

\section{Proof that the generators of ${\cal H}$ commute}
\label{sec-commute}

We can also use table~\ref{fig:numberchart} to help us understand what 
the generators $M_1, \ldots M_a$ of ${\cal H}$ look like.  $M_1$ is always the 
product of all $n\ X_i$'s, and $M_2$ is always the products of all the 
$Z_i$'s.  The other generators are bit more complicated, but still behave 
systematically.  As we advance $i$, they cycle through the sequence 
$I \rightarrow Z \rightarrow X \rightarrow Y$, with a change every
$2^{j-(r-2)}$~qubits for generator $M_r$.  In addition, the NOT switches 
$I \leftrightarrow X$ and $Z \leftrightarrow Y$ whenever it applies --- odd
qubits for even $j$; odd qubits for the first half and even qubits for the
second half for odd $j$.  This immediately implies that every $M_r$ for $r>2$
has equal numbers of $X$'s, $Y$'s, $Z$'s, and $I$'s, namely, $2^{j-2}$ of each. 
Since $j \geq 3$, this means there are an even number of $Y$'s, so $M_r^2 = +1$.

Now, do the generators commute?  Any time two generators have non-trivial 
but different operations on a qubit, we get a factor of -1 when we 
commute them.  Therefore, we can determine if $M_r$ and $M_s$ 
commute by counting the qubits on which they differ and neither is the 
identity.  If this count is even, they commute; if it is odd, they do not.

Since $M_1$ is all $X$'s, it disagrees with $M_r$ (for $r \geq 3$) whenever 
$M_r$ has a $Y$ or a $Z$.  $M_r$ has $2^{j-2}$ of each, so we get $2^{j-1}$ 
factors of -1, and $[M_1, M_r] = 0$.  Similarly, $M_2$ disagrees with $M_r$ 
on $X$'s and $Y$'s, producing $2^{j-2} + 2^{j-2}$ factors of -1, and $[M_2, 
M_r] = 0$ also.  $M_1$ and $M_2$ disagree on every qubit, and since there 
are an even number of qubits, $[M_1, M_2] = 0$.

For $r, s \geq 3$, both $M_r$ and $M_s$ follow the pattern described 
above.  I will consider the cases $s=r+1$, $s=r+2$, and $s>r+2$.  
Table~\ref{fig:commutations} compares $M_r$ and $M_s$ on blocks of size 
$2^{j-(s-2)}$.
\begin{table}
\begin{tabular}{lcccccccccccccccc}
\multicolumn{17}{c}{$r=s+1$:} \\
$M_r$ norm  & I & & I & & Z & & Z & & X & & X & & Y & & Y \\
$M_s$ norm  & I & & Z & & X & & Y & & I & & Z & & X & & Y \\
\\
$M_r$ rev   & X & & X & & Y & & Y & & I & & I & & Z & & Z \\
$M_s$ rev   & X & & Y & & I & & Z & & X & & Y & & I & & Z \\
\\
\multicolumn{17}{c}{$r=s+2$:} \\
$M_r$ norm & I & I & I & I & Z & Z & Z & Z & X & X & X & X & Y & Y & Y & 
Y \\
$M_s$ norm & I & Z & X & Y & I & Z & X & Y & I & Z & X & Y & I & Z & X & 
Y \\
\\
$M_r$ rev  & X & X & X & X & Y & Y & Y & Y & I & I & I & I & Z & Z & Z 
& Z \\
$M_s$ rev  & X & Y & I & Z & X & Y & I & Z & X & Y & I & Z & X & Y & I 
& Z \\
\end{tabular}

\caption{Comparisons of $M_r$ and $M_s$ in blocks of size $2^{j-(s-2)}$ when
the normal cycle applies and when it is reversed by a NOT.}
\label{fig:commutations}
\end{table}

In general, half of each block will be normal and half will be reversed by a 
NOT.  Therefore, the number of factors of -1 from commuting $M_r$ and 
$M_s$ will generally be $2^{j-(s-3)}$ times the total number of non-trivial 
disagreements for the normal and reversed rows.  We also need to 
consider a few special cases.  When $r=3$, the generator never reaches the 
second half of the cycle, so we need to count up the disagreements only in 
the first half of the cycle.  When $s=a=j+2$, the block size is 1, so the NOT 
either affects the whole block or it does not affect any of it.  In this case,
we need to count disagreements only on every other block.  For even $j$, count
the normal disagreements on even numbered blocks and the reversed disagreements 
on odd numbered blocks.  For odd $j$, we must count normal 
disagreements on even-numbered blocks in the first half and odd-numbered 
blocks in the second half; count reversed disgreements on odd-numbered 
blocks in the first half and even-numbered blocks in the second 
half.  We must also consider the combined special case of $r=3$, $s=a$.

For $s=r+1$, the general case gives 4~blocks with normal disagreements 
and 2~blocks with reversed disagreements.  When $r=3$, there are 
2~blocks with normal disagreements and 2~blocks with reversed 
disagreements.  When $s=a$, and $j$ is even, there are 2~blocks with 
normal disagreements and 0~blocks with reversed disagreements.  When 
$s=a$ and $j$ is odd, there are also 2~blocks with normal disagreements 
and 0~blocks with reversed disagreements.  Because $a \geq 5$, we do not 
need to consider the combined special case.  Thus, whenever $s=r+1$, there 
are an even number of disagreements and $M_r$ and $M_s$ commute.

For $s=r+2$, the general case gives 6~blocks with normal disagreements 
and 6~blocks with reversed disagreements.  For $r=3$, there are 2~blocks 
with normal disagreements and 4~blocks with reversed disagreements.  
For $s=a$, $j$ even, there are 4~blocks with normal disagreements and 
2~blocks with reversed disagreements.  For $s=a$, $j$ odd, there are 
2~blocks with normal disagreements and 2~blocks with reversed 
disagreements.  For $r=3$, $s=a$, it does not matter if $j$ is even or odd, 
since we only consider the first half.  In this case, there is 1~block with 
a normal disagreement and 1~block with a reversed disagreement.  In all of 
these cases, the {\em total} number of disagreements is even, so for 
$s=r+2$, $[M_r, M_s] = 0$.

For $s>r+2$, generator $M_s$ completes $2^{s-r-2}$~cycles before $M_r$ 
advances to the next step in the cycle.  This means we can just find the 
number of disagreements by multiplying the number of disagreements for 
$s=r+2$ by $2^{s-r-2}$.  We can do this even for the special cases, since the 
cycle repeats after 4~steps, which does not change the parity.  Thus, there 
will always be an even number of disagreements, and all of the generators 
of ${\cal H}$ commute.

\end{document}